\documentclass[conference]{IEEEtran}
\usepackage{hyperref}

\usepackage{booktabs}
\usepackage{multirow}
\usepackage{longtable}

\usepackage[numbers,square]{natbib}
\usepackage{svg}

\ifCLASSINFOpdf
  \PassOptionsToPackage{pdftex}{graphicx}
\else
\fi
\usepackage[cmex10]{amsmath}

\DeclareMathOperator*{\argmin}{argmin}
\usepackage{array}

\usepackage{mdwmath}
\usepackage{mdwtab}

\usepackage{eqparbox}

\usepackage{caption}
\usepackage{stfloats}

\usepackage{url}

\begin{document}
\title{Generating Realistic Multi-Beat ECG Signals}
\author{\IEEEauthorblockN{Paul P\"ohl\IEEEauthorrefmark{1},
Viktor Schlegel\IEEEauthorrefmark{2}\IEEEauthorrefmark{3},
Hao Li\IEEEauthorrefmark{3}, 
Anil Bharath\IEEEauthorrefmark{1}\IEEEauthorrefmark{2}
}
\IEEEauthorblockA{\IEEEauthorrefmark{1}Department of Bioengineering, Imperial College London, United Kingdom}
\IEEEauthorblockA{\IEEEauthorrefmark{2}Imperial Global Singapore, Imperial College London, Singapore}
\IEEEauthorblockA{\IEEEauthorrefmark{3}Department of Computer Science, University of Manchester, United Kingdom}
Contact: \texttt{v.schlegel@imperial.ac.uk}
}


%
\maketitle
\begin{abstract}
Generating synthetic ECG data has numerous applications in healthcare, from educational purposes to simulating scenarios and forecasting trends. While recent diffusion models excel at generating short ECG segments, they struggle with longer sequences needed for many clinical applications. This paper proposes a novel three-layer synthesis framework for generating realistic long-form ECG signals. We first generate high-fidelity single beats using a diffusion model, then synthesize inter-beat features preserving critical temporal dependencies, and finally assemble beats into coherent long sequences using feature-guided matching. Our comprehensive evaluation demonstrates that the resulting synthetic ECGs maintain both beat-level morphological fidelity and clinically relevant inter-beat relationships. In arrhythmia classification tasks, our long-form synthetic ECGs significantly outperform end-to-end long-form ECG generation using the diffusion model, highlighting their potential for increasing utility for downstream applications. The approach enables generation of unprecedented multi-minute ECG sequences while preserving essential diagnostic characteristics.




\end{abstract}

\section{Introduction}
Generating synthetic time-series data has gained significant attention due to its diverse applications, ranging from augmenting real-world datasets to protecting sensitive information \cite{li2025bridge}. This is particularly crucial in the medical domain, where data privacy regulations, such as GDPR \cite{DBLP:journals/information/GeorgiouL20}, impose strict limitations on the sharing and utilization of patient records \cite{DBLP:conf/bionlp/LiWSBNKZB0N23,DBLP:conf/clef/SchlegelLW0NKBZ23}. The ability to generate high-fidelity synthetic data allows researchers and practitioners to develop and validate machine learning models without compromising patient confidentiality. Moreover, medical datasets are often small, imbalanced, or incomplete, making synthetic data an essential tool for enhancing model robustness and improving generalization in clinical applications \cite{DBLP:journals/corr/abs-2410-03794}.
Numerous generative models, including GANs, VAEs, and most recently \emph{diffusion models}, have shown success in reproducing short segments of physiological signals like electrocardiograms (ECGs) \cite{DBLP:conf/icml/ThapaHKMGM024}. However, when the target application requires \emph{long-sequence} time-series, many models exhibit diminishing performance, resulting in inconsistent or unrealistic transitions over extended durations \cite{DBLP:journals/corr/abs-2408-12249}. In tasks such as arrhythmia classification, where inter-beat intervals (R--R timing) and other multi-beat features play a vital diagnostic role, short synthetic snippets risk omitting crucial temporal dependencies, limiting their real-world utility \cite{DBLP:journals/fdata/KaushikCSDNPD20}.


This paper aims to address the limitations of short-sequence ECG generation by introducing a methodology that synthesizes \emph{long-form} ECG data. We leverage a high-fidelity diffusion model to produce individual beats, then assemble these beats into multi-beat sequences using feature-based alignment and concatenation. Our approach seeks to preserve not only local morphology (e.g., sharp $QRS$ complexes) but also the inter-beat cues crucial for robust downstream analysis. In addition, we propose a structured evaluation pipeline that encompasses both standard time-series metrics and domain-specific criteria such as long-range temporal coherence. By demonstrating improvements in arrhythmia detection tasks when training on these longer synthetic sequences, we highlight the potential for diffusion-based time-series generation to advance research in generating long-sequence synthetic time-series.

\section{Related Work}
Early attempts at synthetic ECG generation primarily focused on short windows, often a single heartbeat or a few consecutive beats. Generative Adversarial Networks (GANs) have been explored extensively for generating time-series and also specifically ECG~\cite{esteban2017realvaluedmedicaltimeseries,Yoon2019,delaney2019synthesisrealisticecgusing,Zhu2019} . Nonetheless, these works reported significant limitations, including difficulties capturing longer-range dynamics, training instability and mode collapse. \citet{Berger2023} reviews the recent developments of GANs in ECG synthesis,  specifically highlighting their inability to generate continuous time-series. Only a few studies, such as \citet{Wulan2020}, have successfully generated extended ECG signals (up to 20 seconds), but report significant downsides, taking months to compute on GPU clusters \cite{Wulan2020}. This underscores the need to develop approaches  capable of synthesizing realistic long-form ECG data.

Variational Auto-encoders (VAEs), while more stable, tend to blur sharp ECG features \cite{Nishikimi2024} over extended segments, diluting clinically relevant markers such as ST-segment deviations.

More recent work has shifted toward \emph{diffusion models}, which learn to de-noise samples to generate data points that achieve higher fidelity and reduced mode collapse \cite{Sikder2023,yuan2023ehrdiff,DBLP:journals/corr/abs-2302-04355,adib2023} Though often applied to image or short-signal generation, several adaptations, including DiffECG~\cite{DBLP:conf/sera/NeifarBMJ24} and DSAT-ECG~\cite{Zama2023}, have demonstrated improved performance on short ECG beats.
Still, these approaches typically generate limited sequences---under a few hundred samples, i.e., a single heart beat---and do not explicitly target multi-beat or multi-lead ECG data that require longer temporal spans.

Parallel to model development, it is crucial to accurately evaluate and assess generated time-series data. \emph{Fidelity} (i.e., realism of generated data) is performed via distribution-level comparisons (such as maximum mean discrepancy) and local shape measures (e.g., dynamic time warping) are used in several studies~\cite{adib2023,nikitin2023}, but longer sequences demand additional checks for continuity and cyclical consistency, which are not easily captured by short-sequence metrics. For downstream \emph{utility}, (i.e., how useful the data would be for downstream tasks, such as ML model training) the ``Train on Synthetic, Test on Real'' (TSTR) protocol is used in many studies~\cite{goodfellow2014gan,Yoon2019,esteban2017realvaluedmedicaltimeseries}.
Although there has been progress towards  unified evaluation frameworks for synthetic data~\cite{nikitin2023,ChinCheong2023}, to date, there is no consensus for a domain-aware evaluation scheme specifically for synthetic ECG data. This is particularly true for longer sequences, which require advanced time-dependent evaluation.

This paper extends the diffusion-based paradigm to long-form ECG generation by proposing a \emph{novel generation methodology}, which integrates an independent feature-generation step and a beat-concatenation process designed to preserve critical inter-beat dependencies. Our proposed pipeline includes a \emph{tiered evaluation}: from beat-level morphology through long-range temporal coherence to classification-based downstream tests. By combining these layers in a unified approach, we aim to better capture the \emph{long-sequence dependencies} that are often overlooked in prior work, and ultimately facilitate more clinically relevant synthetic ECG data.

\section{Methodology}
\begin{figure}[th]
    \centering
    \includegraphics[clip, trim=0.7cm 4.7cm 0.29cm 0cm, width=1\columnwidth]{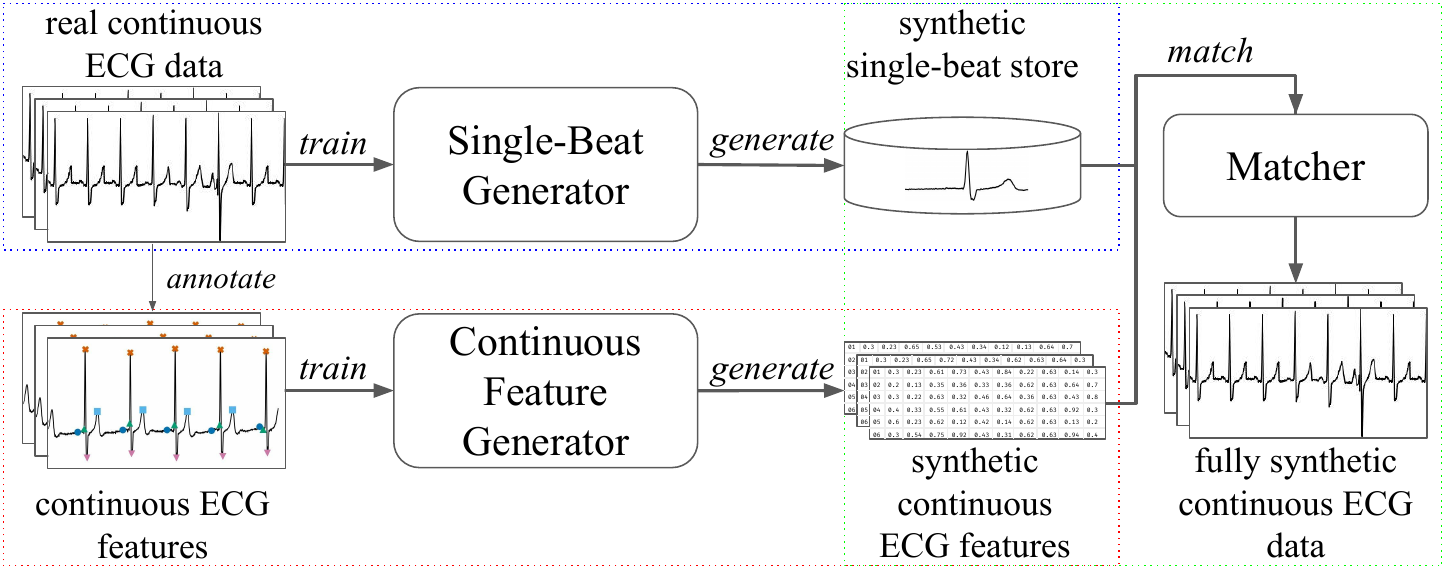}
    \caption{High-level methodology overview: {\color{blue}\emph{(a)}} Beat-level ECG generation using a diffusion model; {\color{red}\emph{(b)}} Feature extraction and generation of synthetic features; {\color{green}\emph{(c)}} Long-sequence assembly via feature-beat matching.}
    \label{fig:method_flowchart}
\end{figure}

We propose a three-layer synthesis framework for generating \emph{long-form} ECG signals: We first learn to generate high-quality beats with a state-of-the-art diffusion model $\mathcal{S}$ (Section A). We then extract advanced per-beat and inter-beat features from long-form ECG signals (such as R-R intervals) and learn to synthesise these with a separate time series generation model $\mathcal{L}$ (Section B). During inference, we oversample $S$ to create a store of high-fidelity single-beat signals. We then sample from $\mathcal{L}$ to obtain a ``scaffold'' to describe the high-level features of the ECG signal and use a matcher $\mathcal{M}$ to match individual beats to it, minimising a cost function (Section C). 
Figure~\ref{fig:method_flowchart} provides an overview.

\subsection{Beat-Level Diffusion Generation}
We use the MIT-BIH Supraventricular Arrhythmia Database\footnote{https://physionet.org/content/svdb/1.0.0/} as our original ECG data, which contains 87 hours of ECG recordings from 47 subjects. For simplicity, we re-assign the present beat-level labels as normal or abnormal (i.e., exhibiting some form of arrhythmia). ECG recordings are complex, high-resolution waveforms, and require a model that can reproduce subtle ECG features, specifically $P$, $Q$, $R$, $S$ and $T$ peaks \cite{adib2023,Zama2023,DBLP:conf/sera/NeifarBMJ24}. Training generative models directly on long sequences often blurs these details or introduces mode collapse~\cite{Berger2023}. Consequently, we first approach the problem of generating high-quality \emph{single-beat} segments: Each segment covers one ECG cycle (approximately $100$ samples, centred around the R-peak). We then use a diffusion model to learn the distribution of these short segments, enabling the model to capture essential morphological nuances while avoiding complexity from extended time dependencies. 
To achieve this, we leverage a set of learning bases that serve as basic elements to construct soft prompts \cite{li2025bridge}. These bases encapsulate intrinsic temporal patterns, including trends, seasonalities, and semantic information pertinent to peaks.
Sampling from this diffusion model yields a large set of synthetic beats (both normal and abnormal) 
each preserving local amplitude and phase characteristics.

\subsection{Multivariate Feature Generation}
\paragraph{Feature Extraction}
Although high-fidelity single beats are essential, real ECG signals exhibit critical \emph{inter-beat} patterns, including varying intervals between $R$ peaks ($R$--$R $intervals, baseline drifts, and amplitude modulations~\cite{Berger2023}. Capturing these longer-term dependencies is necessary for tasks like arrhythmia detection, where multi-beat rhythm properties are central~\cite{Berger2023}. To represent these dependencies, we extract the peaks ($Q$,$R$,$S$,$T$,$P$) for each beat from the real ECG data (e.g., using NeuroKit2~\cite{Makowski2021neurokit} or a similar library). This yields their locations and amplitudes, which we further process to obtain higher level features, such as intervals, focusing on features which have been identified as critical in ECG signals and arrhythmia classification~\cite{vandeleur2022}. Each row in the resulting \emph{feature matrix} corresponds to a single beat, storing both intra-beat (amplitudes, wave durations) and inter-beat (interval to next R-peak) descriptors, indicating how each beat transitions into the next.

\paragraph{Feature Synthesis}
To generate fully synthetic ECG beats rather than reusing the extracted features, we train a \emph{multivariate} model 
on the feature matrix to learn the complex correlations between these descriptors. Sampling from this model yields synthetic feature sequences that capture how features such as $R$--$R$ intervals or baseline amplitude evolve over subsequent beats. The advantage is twofold: \emph{(1)} We preserve the inherent variability and cross-feature relationships observed in real extended ECGs, without using the real, potentially sensitive patient values; and \emph{(2)} W are thus able to generate long feature trajectories (See Appendix, Table~\ref{tab:ramatch_workflow}).


\subsection{Long-Sequence Assembly via Feature Matching}
\paragraph{Matching Algorithm}
Once we have (a) a large set of synthetic beats from the diffusion model (we used 10,000 different beats for normal and abnormal each) and (b) a synthesized sequence of beat feature vectors, we construct extended ECG signals by matching each feature vector with the best-fitting beat. 
Specifically, for each feature vector $\mathbf{f}_n$ 
we select a beat $\mathbf{b}_m$ from the synthetic beat data $\mathcal{B}$ with minimal 
deviations from \(\mathbf{f}_n\): 

\begin{equation}
\label{eq:matching_cost}
    \text{Match}(\mathbf{f}_n, \mathcal{B}) = \argmin_{m \in \{1\ldots|\mathcal{B}|\}} \sum_{k=1}^{d} w_k \left( \mathbf{f}_{n,k} - g_k(\mathbf{b}_m) \right)^2
\end{equation}

where \(f_{n,k}\) is the \(k\)-th feature in \(\mathbf{f}_n\), \(g_k(\mathbf{b}_m)\) extracts the corresponding descriptor (amplitude or interval) from beat \(\mathbf{b}_m\), and \(w_k\) are optional weights for different feature dimensions.

\paragraph{Concatenation and Alignment}
After identifying the best match for each feature vector, we ``stitch'' the selected beats in sequence: $\mathbf{x}_{\text{long}} = \mathbf{b}_{m_1} \Vert \mathbf{b}_{m_2} \Vert \cdots \Vert \mathbf{b}_{m_N}$. We align the R-peak timing and adjust overlaps or gaps to respect the target $R$--$R$ interval. Minor smoothing (e.g., cross-fading or boundary averaging) mitigates abrupt transitions. 


\section{Evaluation and Results}
\label{sec:evaluation}

Our evaluation is organized into three sections: (1) synthetic individual-beats, (2) synthetic inter-beat intervals and amplitudes, and (3) synthetic long-form ECG signals that integrate both local waveforms and inter-beat dependencies. This section outlines our results and discusses both quantitative metrics and qualitative observations.

\subsection{Evaluation of Single-Beat Generation}
The first stage of our pipeline focuses on generating high-fidelity ECG beats with a diffusion model. Each beat, typically spanning $100$--$200$ time steps (at a sampling rate of $128$Hz), is centred around its $R$-peak to preserve the sharp $QRS$ complex and associated waveforms \cite{DBLP:conf/nips/Bedin0DDM24}. We measure morphological fidelity using statistical metrics to quantify alignment and peak amplitude errors to ensure that synthetic beats resemble real ECG morphology \ref{tab:ecg_metrics}. Additionally, visual inspections, including overlay plots of synthetic and real beats, highlight that the diffusion model preserves major ECG landmarks ($P$/$QRS$/$T$ waves), thus preventing mode collapse or excessive smoothing. 

\begin{figure}[ht]
    \centering
    \includesvg[width=0.85\columnwidth]{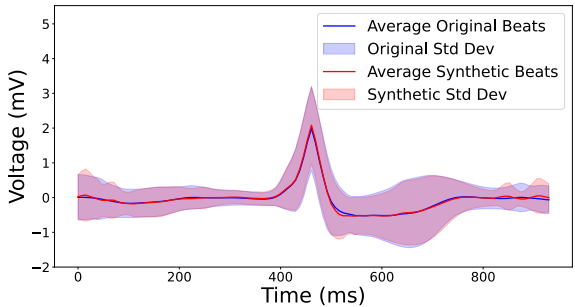}
    \caption{A comparison of the mean of individual generated beats vs original beats with their standard deviations.}
    \label{fig:single_beat_results}
\end{figure}

\subsection{Distribution Analysis of Synthetic ECG Beats}
\begin{figure}[ht]
    \centering
    \includesvg[width=0.85\columnwidth]{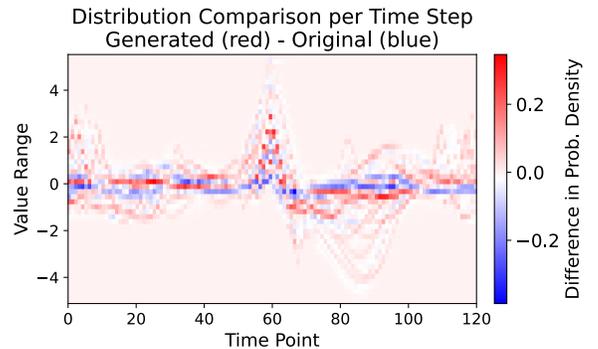}
    \caption{Heatmap of the difference in probability density (Generated - Original) at each time step. Red regions indicate higher synthetic density, while blue regions indicate higher original density.}
    \label{fig:heatmap_diff}
\end{figure}

While our diffusion model successfully preserves the overall morphology---as indicated by similar mean and standard deviation metrics (shown Table~\ref{tab:ecg_metrics} in the Appendix)---the finer statistical properties of the synthetic ECG beats might still differ from those of the real beats. To further analyse this, we compute probability densities of the synthetic and real data for each time step  and visualize their differences as a heat map in Figure~\ref{fig:heatmap_diff}.
The visualization reveals that, despite the close match in aggregate statistics, there are still subtle differences in the underlying distributions at various time steps, especially at time steps with a large standard deviation. Overall, the divergence of the distributions per timestep is moderately low ($0.096$, see Table~\ref{tab:ecg_metrics} in Appendix).

\subsection{Arrhythmia Classification Using Individual Beats}

We further assess the viability of our synthetic single-beat waveforms for arrhythmia classification using a Train on Synthetic, Test on Real (TSTR) framework. In parallel, classifiers are also trained on original individual beats to provide a performance baseline. Table~\ref{tab:single_beat_metrics} summarizes the results for several classifiers—including Balanced SVM, Decision Tree, Naive Bayes, and SVM—evaluated on both synthetic and original data. Notably, models trained on synthetic beats achieve competitive performance, with SVM-based approaches demonstrating robust accuracy, precision, and recall in distinguishing normal from arrhythmic beats. These promising results validate the utility of our synthetic data for arrhythmia classification, even though multi-beat temporal context is not incorporated.

\begin{table}[ht]
\centering
\caption{Arrhythmia classification performance using individual beats. Results for both synthetic (TSTR) and original data are shown.}
\label{tab:single_beat_metrics}
\begin{tabular}{lcccccc}
\toprule
\multirow{2}{*}{Model} & \multicolumn{2}{c}{Acc} & \multicolumn{2}{c}{Pr.} & \multicolumn{2}{c}{Rec.} \\
\cmidrule{2-3}\cmidrule{4-5}\cmidrule{6-7}
  & (N) & (A) & (N) & (A) & (N) & (A) \\
\midrule
\multicolumn{7}{l}{\textbf{Original Data}} \\
Balanced SVM   & 0.97 & 0.83 & 0.92 & 0.92 & 0.97 & 0.83 \\
Decision Tree  & 0.89 & 0.86 & 0.93 & 0.80 & 0.89 & 0.86 \\
Naive Bayes    & 0.89 & 0.45 & 0.77 & 0.68 & 0.89 & 0.45 \\
SVM            & 0.98 & 0.79 & 0.91 & 0.96 & 0.98 & 0.79 \\
\midrule
\multicolumn{7}{l}{\textbf{Synthetic Data}} \\
Balanced SVM   & 0.84 & 0.76 & 0.88 & 0.70 & 0.84 & 0.76 \\
Decision Tree  & 0.77 & 0.77 & 0.87 & 0.62 & 0.77 & 0.77 \\
Naive Bayes    & 0.72 & 0.55 & 0.77 & 0.49 & 0.72 & 0.55 \\
SVM            & 0.84 & 0.76 & 0.88 & 0.70 & 0.84 & 0.76 \\
\bottomrule
\end{tabular}
\end{table}

\subsection{Evaluation of Feature Extraction and Generation}
\begin{figure}[t]
    \centering
    \includesvg[width=0.95\columnwidth]{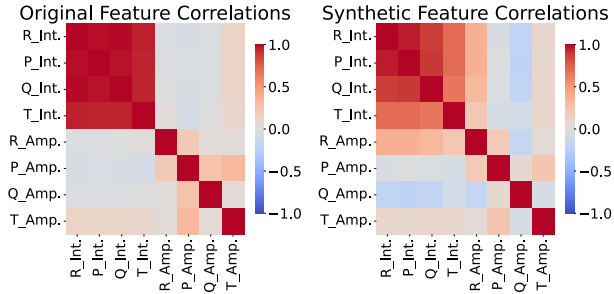}
    \caption{Overview of pairwise correlations of features from the original and synthetic ECG data. The synthetic data largely preserves important correlations.}
    \label{fig:feature_extraction_results}
\end{figure}

While single-beat fidelity is critical, many cardiac conditions manifest through inter-beat variability \cite{DBLP:conf/cinc/MariaBCGLPMCGPVP22}. 
In Figure~\ref{fig:feature_extraction_results} we visualise the quality of generated synthetic features by measuring their inter-correlation and comparing to those extracted from real data. We further measure the distributional differences of these features in Table~\ref{tab:feature_metrics_updated}.
\begin{table}[ht]
\centering
\caption{Advanced metrics for feature generation, including divergence measures (KL), distributional distances (MMD and Wasserstein), the Kolmogorov-Smirnov (KS) test statistic, and difference in Mean.}
\label{tab:feature_metrics_updated}
\resizebox{.95\columnwidth}{!}{%
\begin{tabular}{lccccccccc}
\toprule
\textbf{Feature} & \textbf{KL Div.} & \textbf{MMD} & \textbf{Wass. Dist.} & \textbf{KS} & \textbf{Mean Diff.} \\
\midrule
$R_{Int}$ & 0.038 & 0.045 & 1.967 & 0.043 &  0.522 \\
$P_{Int}$ & 0.023 & 0.044 & 2.088 & 0.040 & 0.640 \\
$Q_{Int}$ & 0.021 & 0.045 & 2.676 & 0.053 &  0.670 \\
$T_{Int}$ & 0.037 & 0.137 & 4.708 & 0.097 &  0.280 \\
$R_{Amp}$ & 0.250 & 0.069 & 0.121 & 0.155 & 0.034 \\
$P_{Amp}$ & 0.125 & 0.017 & 0.046 & 0.052 &  0.038 \\
$Q_{Amp}$ & 0.032 & 0.008 & 0.030 & 0.090 & 0.002 \\
$T_{Amp}$ & 0.139 & 0.017 & 0.069 & 0.125 & 0.023 \\
\bottomrule
\end{tabular}
}

\end{table}

The combination of visual insights from Figure~\ref{fig:feature_extraction_results} and the quantitative analysis in Table~\ref{tab:feature_metrics_updated} underscores the ability of our feature-level generation model to capture both local and global dynamics of ECG signals. While interval features ($\cdot_{Int}$) display low divergence and distance values—indicating that their synthetic distributions closely mirror those of the real data—amplitude features (especially $R_{Amp}$) exhibit higher divergence measures, highlighting the increased challenge in accurately replicating these aspects. 
These findings are largely in line with the literature~\cite{hollmann2025accurate}, thus we are confident that the generation of the features and inclusion of larger and more complex features is feasible, further improving the generation process. 


\begin{figure}[ht]
    \centering
    \includegraphics[width=0.8\linewidth]{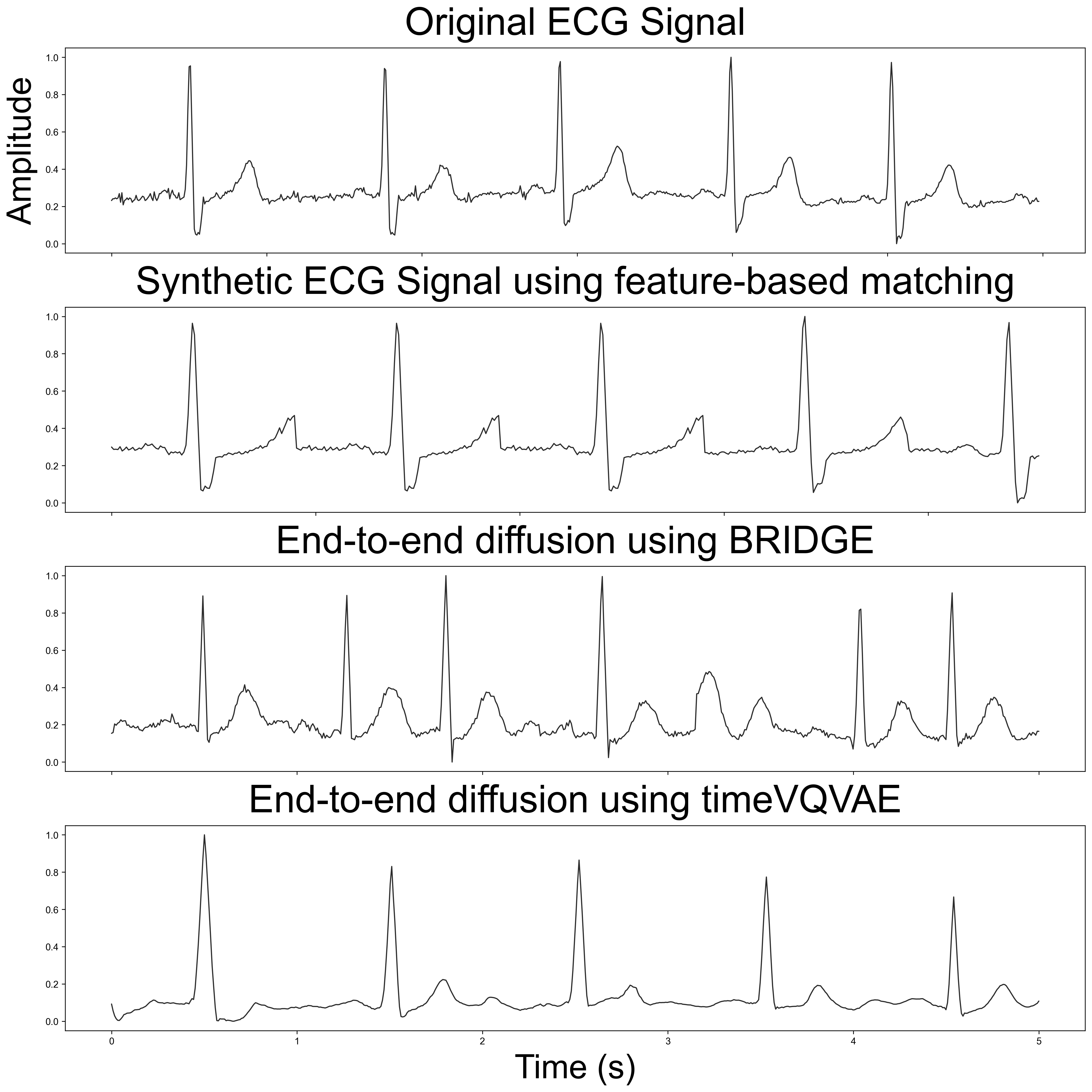}
    \caption{Comparison of randomly chosen long-form ECGs. We compare an original signal against a synthetic signal using our approach and 2 signal generated End-to-end using the BRIDGE \cite{li2025bridge} diffusion model and timeVQVAE \cite{lee2023vectorquantizedtimeseries}(bottom).}
    \label{fig:long_form_results}
\end{figure}
\subsection{Evaluation of Long-Form ECG Assembly}

Qualitative inspection of concatenated signals indicates that R-peaks are appropriately aligned with minimal artifacts \ref{fig:long_form_results}, while quantitative checks confirm that the distribution of R-R intervals in the assembled ECG preserves the synthetic feature trajectory (as can be seen in the Appendix Table~\ref{tab:ramatch_summary}). Such alignment suggests that both short-range and long-range ECG characteristics are realistically captured.

Finally, to contextualise the framework with existing literature, we compare downstream performance of classifiers trained on generated multi-beat ECG generated by our proposed framework and an end-to-end diffusion baseline. We limit the number of beats to 5, as more beats lead to OOM errors for the end-to-end approach. Results are summarised in Table~\ref{tab:long-ecg-comparison}. Clearly, our proposed approach outperforms the end-to-end synthesis and comes close to the performance of classifiers trained on the original data.

\begin{table}[ht]
\centering
\caption{Arrhythmia classification performance on original, synthetic, and End-to-end generated data, averaged across four classifiers.}
\label{tab:long-ecg-comparison}
    \begin{tabular}{lcccccc}
    \toprule
    \multirow{2}{*}{Model} & \multicolumn{2}{c}{Acc} & \multicolumn{2}{c}{Pr.} & \multicolumn{2}{c}{F1} \\
    \cmidrule(lr){2-3}\cmidrule(lr){4-5}\cmidrule(lr){6-7}
      & (N) & (A) & (N) & (A) & (N) & (A) \\
    \midrule
    Original                    & 0.933 & 0.735 & 0.881 & 0.838 & 0.906 & 0.778 \\
    Synthetic feature-based          & 0.948 & 0.526 & 0.964 & 0.474 & 0.956 & 0.479 \\
    BRIDGE End-to-end       & 0.283 & 0.764 & 0.507 & 0.443 & 0.222 & 0.470 \\
    TimeVQVAE End-to-end        & 0.764 & 0.250 & 0.719 & 0.439 & 0.662 & 0.146 \\

    \bottomrule
    \end{tabular}
\end{table}

\section{Conclusion}
We present a synthetic ECG generation approach that addresses a critical problem by providing a scalable framework for creating clinically meaningful long-form signals. By decomposing the process into beat morphology, temporal patterns, and sequence assembly, we overcome limitations of existing methods that struggle with extended time dependencies. This enables scaling to multiple minutes of continuous ECG while maintaining both local and global characteristics.

Multi-tiered evaluation demonstrates our framework successfully synthesizes ECG data that \emph{(i)} preserves high-resolution beat morphology, \emph{(ii)} accurately reproduces beat-to-beat variability through feature-based modelling, and \emph{(iii)} assembles coherent extended signals via feature-guided concatenation.

This comprehensive approach outperforms end-to-end baselines and holds promise for clinical applications. Future improvements could include additional synthetic features, enhanced matching functions, and better filtering for higher quality single beats.

\section*{Acknowledgement}
This research is part of the IN-CYPHER programme and is supported by the National Research Foundation, Prime Minister’s Office, Singapore under its Campus for Research Excellence and Technological Enterprise (CREATE) programme. We are grateful for the support provided by Research IT to him in form of access to the Computational Shared Facility at The University of Manchester and the computational facilities at the Imperial College Research Computing Service\footnote{DOI: \url{https://doi.org/10.14469/hpc/2232}}.

\bibliography{refs,references}
\bibliographystyle{IEEEtranN}



\appendix
\section{Example Appendix}
\label{sec:appendix}

\subsection{Single Beat Generation Analysis}
\begin{figure}[ht]
    \centering
    \includesvg[width=1\columnwidth]{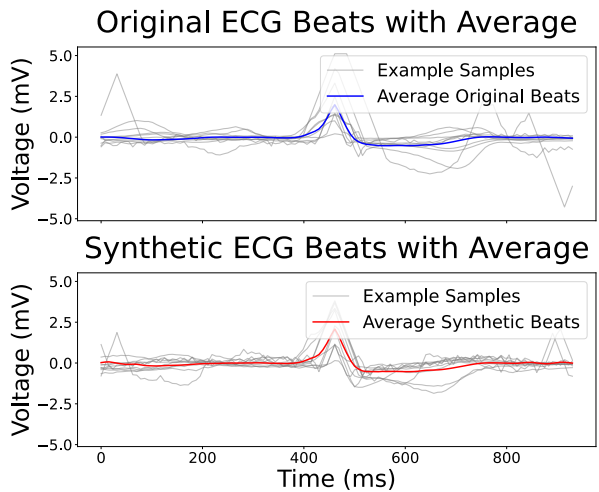}
    \caption{Examples of individual beats and their averages, showing (a) real ECG waveforms, (b) synthetic waveforms}
    \label{fig:single_beat_results2}
\end{figure}

The synthesis framework relies on generating high-fidelity individual ECG beats, which are subsequently matched. Generating a large variety of high quality, accurate synthetic beats is crucial in allowing the methodology to produce an accurate final long-form signal. Figure~\ref{fig:single_beat_results2} compares real and synthetic beat morphology, highlighting the similarity of the synthetic beats. This further validates our diffusion model, successfully being able to capture essential electrophysiological patterns described in Section~III-A. The grey lines, showing 10 individual beats for each, indicate effective learning of both typical beat characteristics and natural variability.

\subsection{Quantitative Beat Quality Metrics}
Qualitative evaluation of the beats confirms visually the quality of the synthetic beats. This is confirmed in Table~\ref{tab:ecg_metrics}, which provides a comprehensive statistical validation of synthetic beat quality. The DTW distance of 2.06717 and Fréchet distance of 0.09619 suggest strong temporal and structural similarity to real beats. Low RMSE (0.03368) and MAE (0.02392) values indicate precise amplitude matching, crucial for clinical relevance. The JS divergence of 0.01177 and Wasserstein distance of 0.05120 confirm distributional alignment at both global and local waveform scales. These metrics quantitatively validate the the single beats quality, as described in Section~III-A.

\begin{table}[ht]
    \centering
    \caption{Evaluation Metrics for Synthetic individual ECG Beats}
    \begin{tabular}{@{}l r@{}}
        \toprule
        \textbf{Metric} & \textbf{Value} \\
        \midrule
        DTW distance between average ECG beats & 2.06717 \\
        Fréchet distance between average ECG beats & 0.09619 \\
        Euclidean distance between average ECG beats & 0.36893 \\
        RMSE between average ECG beats & 0.03368 \\
        MAE between average ECG beats & 0.02392 \\
        MSE between average ECG beats & 0.00113 \\
        PRD between average ECG beats & 8.531 \\
        Average KL Divergence & 0.09591 \\
        Average JS Divergence & 0.01177 \\
        Average MMD & 0.02492 \\
        Average Wasserstein Distance & 0.05120 \\
        Average KS Statistic & 0.06579 \\
        Average Mean Difference & 0.02392 \\
        Average Variance Difference & 0.04736 \\
        Average Skewness Difference & 0.51102 \\
        \bottomrule
    \end{tabular}
    \label{tab:ecg_metrics}
\end{table}

\subsection{R-Peak Amplitude Matching}
The $R_{Amp}$ matching statistics in Table~\ref{tab:ramatch_summary} showcase the $R_{Amp}$ matching of chosen single beats to the synthetic features, with an example of generating a 100 beat long signal. This confirms our framework's ability to select physiologically plausible beats during long-form assembly, 89\% of beats showing $<0.03$mV absolute difference from target amplitudes. This confirms that the matching algorithm (Section~III-C) works, and is able to  effectively leverages our large synthetic beat repository. 

\begin{table}[ht]
    \centering
    \caption{Summary of $R_{Amp}$ Matching Across Synthetic Beats}
    \label{tab:ramatch_summary}
    \begin{tabular}{@{}l c@{}}
        \toprule
        \textbf{Absolute Difference Range (mV)} & \textbf{Number of Beats} \\
        \midrule
        0.00 -- 0.01 & 29 \\
        0.01 -- 0.02 & 37 \\
        0.02 -- 0.03 & 23 \\
        0.03 -- 0.04 & 4 \\
        0.04 -- 0.05 & 3 \\
        0.05 -- 0.06 & 1 \\
        0.06 -- 0.07 & 1 \\
        $>$ 0.07    & 1 \\
        \midrule
        \textbf{Total Beats} & 100 \\
        \bottomrule
    \end{tabular}
\end{table}

\subsection{Feature Matching Workflow}
Table~\ref{tab:ramatch_workflow} exemplifies our beat selection process during long-form synthesis. With 10,000 candidate beats per selection, the algorithm typically finds matches within 0.02mV of target amplitudes using just 8-13 candidates evaluated. This efficiency enables scalable generation of hour-long ECGs while maintaining an optimal balance between computational cost and physiological accuracy, as discussed in Section~III-C.

\begin{table*}[ht]
    \centering
    \caption{$R_{Amp}$ Matching Workflow for Synthetic Beats}
    \label{tab:ramatch_workflow}
    \begin{tabular}{@{}l c c c c c@{}}
        \toprule
        \textbf{Beat} & \textbf{Target $R_{Amp}$} & \textbf{Best Match $R_{Amp}$} & \textbf{Absolute Diff.} & \textbf{Candidates} & \textbf{Beats to choose from} \\
        \midrule
        0  & 0.30 & 0.35 & 0.04 & 10 & 10\,000 normal beats \\
        1  & 1.42 & 1.41 & 0.00 & 9  & 10\,000 normal beats \\
        2  & 1.35 & 1.37 & 0.02 & 13 & 10\,000 normal beats \\
        3  & 1.43 & 1.40 & 0.02 & 13 & 10\,000 normal beats \\
        4  & 1.41 & 1.40 & 0.00 & 10 & 10\,000 normal beats \\
        5  & 1.38 & 1.40 & 0.03 & 8  & 10\,000 normal beats \\
        6  & 1.50 & 1.50 & 0.01 & 11 & 10\,000 normal beats \\
        7  & 1.40 & 1.38 & 0.02 & 9  & 10\,000 normal beats \\
        8  & 1.45 & 1.47 & 0.02 & 13 & 10\,000 normal beats \\
        9  & 1.40 & 1.40 & 0.01 & 10 & 10\,000 normal beats \\
        ...  & ... & ... & ... & ... & ... \\
        \bottomrule
    \end{tabular}
\end{table*}

\subsection{Downstream Classification Performance}
Table~\ref{tab:long-form-classification} validates the clinical utility of synthetic ECGs through TSTR(Train on Synthetic Test on Real) classifier testing. It significantly manages to outperform the end-to-end diffusion classifcation and is even able to achieve performance close to the original for some models.

\begin{table*}[ht]
\centering
\caption{MIT-BIH Supraventricular Arrhythmia: Comparison of classifiers trained on original vs.\ end-to-end diffused vs.\ Synthetic-Long ECG vs.\ TimeVQVAE end-to-end}
\label{tab:long-form-classification}
\begin{tabular}{@{}l l c c c c c c c c c c c@{}}
\toprule
\textbf{Data} & \textbf{Model} & \textbf{A} & \textbf{Acc. N} & \textbf{Acc. A} &
\textbf{Prec (N)} & \textbf{Prec (A)} & \textbf{Rec (N)} & \textbf{Rec (A)} &
\textbf{F1 (N)} & \textbf{F1 (A)} & \textbf{MCC} \\
\midrule
\multicolumn{12}{@{}l}{\textbf{Balanced SVM Classifier}} \\
\midrule
\textbf{Original}             &
    & 0.9230  & 0.9660  & 0.8340  & 0.9220  & 0.9240  & 0.9660  & 0.8340  & 0.9440  & 0.8760  & 0.8230 \\
\textbf{BRIDGE end-to-end}   &
    & 0.2820  & 0.0433  & 0.8674  & 0.4440  & 0.2700  & 0.0433  & 0.8674  & 0.0790  & 0.4120  & -0.1600 \\
\textbf{Synthetic-long ECG}    &
    & 0.9058 & 0.9298 & 0.5883 & 0.9677 & 0.3874 & 0.9298 & 0.5883 & 0.9483 & 0.4672 & 0.4289 \\
\textbf{TimeVQVAE end-to-end} &
    & 0.7130  & 0.9732  & 0.0767  & 0.7205  & 0.5395  & 0.9732  & 0.0767  & 0.8280  & 0.1344  & 0.1140 \\
\midrule
\multicolumn{12}{@{}l}{\textbf{Decision Tree Classifier}} \\
\midrule
\textbf{Original}             &
    & 0.8820  & 0.8910  & 0.8640  & 0.9300  & 0.7960  & 0.8910  & 0.8640  & 0.9100  & 0.8280  & 0.7410 \\
\textbf{BRIDGE end-to-end}  &
    & 0.3710  & 0.1572  & 0.8943  & 0.7843  & 0.3026  & 0.1572  & 0.8943  & 0.2620  & 0.4522  & 0.0670 \\
\textbf{Synthetic-long ECG}    &
    & 0.8941 & 0.9128 & 0.6467 & 0.9716 & 0.3589 & 0.9128 & 0.6467 & 0.9413 & 0.4616 & 0.4300 \\
\textbf{TimeVQVAE end-to-end} &
    & 0.7071  & 0.9951  & 0.0025  & 0.7093  & 0.1744  & 0.9951  & 0.0025  & 0.8282  & 0.0050  & -0.0165 \\
\midrule
\multicolumn{12}{@{}l}{\textbf{Naive Bayes Classifier}} \\
\midrule
\textbf{Original}             &
    & 0.7480  & 0.8930  & 0.4520  & 0.7690  & 0.6750  & 0.8930  & 0.4520  & 0.8270  & 0.5420  & 0.3920 \\
\textbf{BRIDGE end-to-end}  &
    & 0.5350  & 0.8630  & 0.4530  & 0.2830  & 0.9300  & 0.8630  & 0.4530  & 0.4260  & 0.6090  & 0.2600 \\
\textbf{Synthetic-long ECG}    &
    & 0.9356 & 0.9711 & 0.4658 & 0.9601 & 0.5487 & 0.9711 & 0.4658 & 0.9656 & 0.5039 & 0.4715 \\
\textbf{TimeVQVAE end-to-end} &
    & 0.3312  & 0.0895  & 0.9224  & 0.7384  & 0.2929  & 0.0895  & 0.9224  & 0.1597  & 0.4446  & 0.0193 \\
\midrule
\multicolumn{12}{@{}l}{\textbf{SVM Classifier}} \\
\midrule
\textbf{Original}             &
    & 0.9190  & 0.9820  & 0.7890  & 0.9050  & 0.9560  & 0.9820  & 0.7890  & 0.9420  & 0.8650  & 0.8150 \\
\textbf{BRIDGE end-to-end}  &
    & 0.2930  & 0.0690  & 0.8420  & 0.5160  & 0.2700  & 0.0690  & 0.8420  & 0.1210  & 0.4090  & -0.1380 \\
\textbf{Synthetic-long ECG}    &
    & 0.9392 & 0.9796 & 0.4046 & 0.9561 & 0.5992 & 0.9796 & 0.4046 & 0.9677 & 0.4830 & 0.4618 \\
\textbf{TimeVQVAE end-to-end} &
    & 0.7098  & 1.0000  & 0.0002  & 0.7098  & 0.7500  & 1.0000  & 0.0002  & 0.8303  & 0.0004  & 0.0094 \\
\bottomrule
\end{tabular}
\end{table*}

\subsection{Tested on only MIT-BIH Arrhythmia Database}
Additionally, we tested the workflow on the standard MIT-BIH Arrhythmia Database, which performs better than the more extensive Supraventricular Database. It shows the same result, the synthetic-long ECG format significantly outperforming end-to-end generation.

\begin{table*}[ht]
\centering
\caption{Testing on basic MIT-BIH Arrhythmia Database: Comparison of classifiers trained on original vs.\ end-to-end diffused vs.\ Synthetic-Long ECG}
\label{tab:new-long-form-classification}
\begin{tabular}{@{}l l c c c c c c c c c c@{}}
\toprule
\textbf{Data} & \textbf{Model} & \textbf{A} & \textbf{Acc. N} & \textbf{Acc. A} &
\textbf{Prec (N)} & \textbf{Prec (A)} & \textbf{Rec (N)} & \textbf{Rec (A)} &
\textbf{F1 (N)} & \textbf{F1 (A)} & \textbf{MCC} \\
\midrule

\multicolumn{12}{@{}l}{\textbf{Logistic Regression}} \\
\midrule
\textbf{Original}             & & 0.9572 & 0.9786 & 0.8738 & 0.9680 & 0.9126 & 0.9786 & 0.8738 & 0.9733 & 0.8928 & 0.8664 \\
\textbf{End‐to‐end generation}& & 0.4380 & 0.3241 & 0.8826 & 0.9151 & 0.2505 & 0.3241 & 0.8826 & 0.4787 & 0.3903 & 0.1851 \\
\textbf{Synthetic‐long ECG}    & & 0.9325 & 0.9275 & 0.9519 & 0.9869 & 0.7708 & 0.9275 & 0.9519 & 0.9563 & 0.8518 & 0.8163 \\

\midrule
\multicolumn{12}{@{}l}{\textbf{Balanced SVM Classifier}} \\
\midrule
\textbf{Original}             & & 0.9850 & 0.9860 & 0.9809 & 0.9951 & 0.9473 & 0.9860 & 0.9809 & 0.9905 & 0.9638 & 0.9545 \\
\textbf{End‐to‐end generation}& & 0.6808 & 0.6060 & 0.9730 & 0.9887 & 0.3873 & 0.6060 & 0.9730 & 0.7514 & 0.5541 & 0.4666 \\
\textbf{Synthetic‐long ECG}    & & 0.9728 & 0.9702 & 0.9828 & 0.9955 & 0.8941 & 0.9702 & 0.9828 & 0.9827 & 0.9364 & 0.9208 \\

\midrule
\multicolumn{12}{@{}l}{\textbf{Decision Tree Classifier}} \\
\midrule
\textbf{Original}             & & 0.9773 & 0.9858 & 0.9440 & 0.9857 & 0.9446 & 0.9858 & 0.9440 & 0.9858 & 0.9443 & 0.9301 \\
\textbf{End‐to‐end generation}& & 0.7255 & 0.6649 & 0.9623 & 0.9857 & 0.4237 & 0.6649 & 0.9623 & 0.7941 & 0.5883 & 0.5067 \\
\textbf{Synthetic‐long ECG}    & & 0.9543 & 0.9489 & 0.9757 & 0.9935 & 0.8301 & 0.9489 & 0.9757 & 0.9707 & 0.8970 & 0.8726 \\

\midrule
\multicolumn{12}{@{}l}{\textbf{Naive Bayes Classifier}} \\
\midrule
\textbf{Original}             & & 0.9224 & 0.9619 & 0.7681 & 0.9419 & 0.8376 & 0.9619 & 0.7681 & 0.9518 & 0.8013 & 0.7543 \\
\textbf{End‐to‐end generation}& & 0.6366 & 0.5711 & 0.8922 & 0.9539 & 0.3475 & 0.5711 & 0.8922 & 0.7145 & 0.5002 & 0.3737 \\
\textbf{Synthetic‐long ECG}    & & 0.7937 & 0.7808 & 0.8443 & 0.9514 & 0.4964 & 0.7808 & 0.8443 & 0.8577 & 0.6252 & 0.5291 \\

\midrule
\multicolumn{12}{@{}l}{\textbf{SVM Classifier}} \\
\midrule
\textbf{Original}             & & 0.9864 & 0.9967 & 0.9461 & 0.9864 & 0.9867 & 0.9967 & 0.9461 & 0.9915 & 0.9660 & 0.9578 \\
\textbf{End‐to‐end generation}& & 0.8063 & 0.7650 & 0.9679 & 0.9894 & 0.5132 & 0.7650 & 0.9679 & 0.8628 & 0.6707 & 0.6069 \\
\textbf{Synthetic‐long ECG}    & & 0.9728 & 0.9702 & 0.9828 & 0.9955 & 0.8941 & 0.9702 & 0.9828 & 0.9827 & 0.9364 & 0.9208 \\

\midrule
\multicolumn{12}{@{}l}{\textbf{XGBClassifier}} \\
\midrule
\textbf{Original}             & & 0.9897 & 0.9960 & 0.9652 & 0.9912 & 0.9841 & 0.9960 & 0.9652 & 0.9936 & 0.9746 & 0.9682 \\
\textbf{End‐to‐end generation}& & 0.8485 & 0.8147 & 0.9802 & 0.9938 & 0.5753 & 0.8147 & 0.9802 & 0.8954 & 0.7250 & 0.6726 \\
\textbf{Synthetic‐long ECG}    & & 0.9790 & 0.9772 & 0.9862 & 0.9964 & 0.9171 & 0.9772 & 0.9862 & 0.9867 & 0.9504 & 0.9381 \\

\midrule
\multicolumn{12}{@{}l}{\textbf{Random Forest Classifier}} \\
\midrule
\textbf{Original}             & & 0.9892 & 0.9958 & 0.9631 & 0.9906 & 0.9833 & 0.9958 & 0.9631 & 0.9932 & 0.9731 & 0.9664 \\
\textbf{End‐to‐end generation}& & 0.8783 & 0.8530 & 0.9770 & 0.9932 & 0.6298 & 0.8530 & 0.9770 & 0.9177 & 0.7659 & 0.7190 \\
\textbf{Synthetic‐long ECG}    & & 0.9737 & 0.9694 & 0.9906 & 0.9975 & 0.8923 & 0.9694 & 0.9906 & 0.9833 & 0.9389 & 0.9243 \\

\bottomrule
\end{tabular}
\end{table*}

\end{document}